\documentclass[8.5pt,twoside,twocolumn]{article}
\usepackage[super,sort&compress,comma]{natbib} 
\usepackage{mhchem}
\usepackage{times}
\usepackage{sectsty}
\usepackage{balance} 

\usepackage{graphicx} 
\usepackage{lastpage}
\usepackage[format=plain,justification=raggedright,singlelinecheck=false,font=small,labelfont=bf,labelsep=space]{caption} 
\usepackage{fancyhdr}
\usepackage{amssymb,amsmath,MnSymbol,ifsym,latexsym,bm}
\usepackage[Euler]{upgreek}
\usepackage{bm}
\usepackage{color}

\newcommand{\Oh}{\mathrm{Oh}}
\newcommand{\Bo}{\mathrm{Bo}}
\newcommand{\Pe}{\mathrm{Pe}}
\newcommand{\etas}{\eta_s}

\newcommand{\eeta}{\overline{\eta}}
\newcommand{\eetap}{\overline{\eta}_\mathrm{p}}
\newcommand{\eetarel}{\eeta_\mathrm{rel} }

\newcommand{\ieeta}{[\eeta]}
\newcommand{\edot}{\dot{\epsilon}}
\newcommand{\edota}{\dot{\epsilon}_\mathrm{avg}}
\newcommand{\tauR}{\tau_\mathrm{R}}

\newcommand{\kBT}{k_\mathrm{B} T}
\newcommand{\kB}{k_\mathrm{B}}

\newcommand{\zh}{\zeta_h}
\newcommand{\zt}{\zeta_t}

\newcommand{\bom}{\bm{\Upomega}}

\newcommand{\dA}{\widetilde{\chi}}
\newcommand{\dB}{\widetilde{\beta}}
\newcommand{\dG}{\widetilde{\gamma}}
\newcommand{\dS}{\widetilde{\sigma}}

\newcommand{\ut}{\bm{\updelta}}

\renewcommand{\vec}[1]{\bm{\mathrm{#1}}}

\begin{document}

\thispagestyle{plain}
\fancypagestyle{plain}{
\fancyhead[L]{\includegraphics[height=8pt]{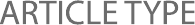}}
\fancyhead[C]{\hspace{-1cm}\includegraphics[height=20pt]{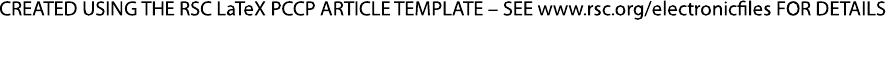}}
\fancyhead[R]{\includegraphics[height=10pt]{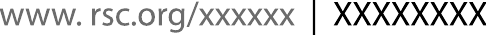}\vspace{-0.2cm}}
\renewcommand{\headrulewidth}{1pt}}
\renewcommand{\thefootnote}{\fnsymbol{footnote}}
\renewcommand\footnoterule{\vspace*{1pt}%
\hrule width 3.4in height 0.4pt \vspace*{5pt}} 
\setcounter{secnumdepth}{5}

\makeatletter 
\def\subsubsection{\@startsection{subsubsection}{3}{10pt}{-1.25ex plus -1ex minus -.1ex}{0ex plus 0ex}{\normalsize\bf}} 
\def\paragraph{\@startsection{paragraph}{4}{10pt}{-1.25ex plus -1ex minus -.1ex}{0ex plus 0ex}{\normalsize\textit}} 
\renewcommand\@biblabel[1]{#1}            
\renewcommand\@makefntext[1]%
{\noindent\makebox[0pt][r]{\@thefnmark\,}#1}
\makeatother 
\renewcommand{\figurename}{\small{Fig.}~}
\sectionfont{\large}
\subsectionfont{\normalsize} 

\fancyfoot{}
\fancyfoot[LO,RE]{\vspace{-7pt}\includegraphics[height=9pt]{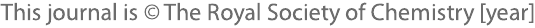}}
\fancyfoot[CO]{\vspace{-7.2pt}\hspace{12.2cm}\includegraphics{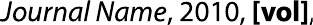}}
\fancyfoot[CE]{\vspace{-7.5pt}\hspace{-13.5cm}\includegraphics{RF}}
\fancyfoot[RO]{\footnotesize{\sffamily{1--\pageref{LastPage} ~\textbar  \hspace{2pt}\thepage}}}
\fancyfoot[LE]{\footnotesize{\sffamily{\thepage~\textbar\hspace{3.45cm} 1--\pageref{LastPage}}}}
\fancyhead{}
\renewcommand{\headrulewidth}{1pt} 
\renewcommand{\footrulewidth}{1pt}
\setlength{\arrayrulewidth}{1pt}
\setlength{\columnsep}{6.5mm}
\setlength\bibsep{1pt}

\twocolumn[
  \begin{@twocolumnfalse}
\noindent\LARGE{\textbf{Motility induced changes in viscosity of suspensions of swimming microbes in extensional flows$^\dag$}}
\vspace{0.6cm}

\noindent\large{\textbf{Amarin G. McDonnell,\textit{$^{a}$} Tilvawala C. Gopesh,\textit{$^{a}$} Jennifer Lo,\textit{$^{b}$}
Moira O'Bryan,\textit{$^{b}$} Leslie Y. Yeo,\textit{$^{c}$} James R. Friend,\textit{$^{c\ddag}$} and Ranganathan Prabhakar$^{\ast}$\textit{$^{a}$}}}\vspace{0.5cm}

\noindent\textit{\small{\textbf{Received Xth XXXXXXXXXX 20XX, Accepted Xth XXXXXXXXX 20XX\newline
First published on the web Xth XXXXXXXXXX 200X}}}

\noindent \textbf{\small{DOI: 10.1039/b000000x}}
\vspace{0.6cm}

\noindent \normalsize{Suspensions of motile cells are model systems for understanding the unique mechanical properties of living materials which often consist of ensembles of self-propelled particles. We present here a quantitative comparison of theory against experiment for the rheology of such suspensions. The influence of motility on viscosities of cell suspensions is studied using a novel acoustically-driven microfluidic capillary-breakup extensional rheometer. Motility increases the extensional viscosity of suspensions of algal pullers, but decreases it in the case of bacterial or sperm pushers. A recent model [Saintillan, \textit{Phys. Rev. E},  2010, \textbf{81}, 56307] for dilute active suspensions is extended to obtain predictions for higher concentrations, after independently obtaining parameters such as swimming speeds and diffusivities. We show that details of body and flagellar shape can significantly determine macroscale rheological behaviour. }
\vspace{0.5cm}
 \end{@twocolumnfalse}
  ]

\section{Introduction}
\footnotetext{\dag~Electronic Supplementary Information (ESI) available. See DOI: 10.1039/b000000x/}


\footnotetext{\textit{$^{a}$~Department of Mechanical and Aerospace Engineering, Monash University, Clayton, Australia. Fax: 613 9905 1825; Tel: 613 9905 3480; E-mail: prabhakar.ranganathan@monash.edu}}
\footnotetext{\textit{$^{b}$~Department of Anatomy and Developmental Biology, Monash University, Clayton, Australia. }}
\footnotetext{\textit{$^{c}$~Micro/Nanophysics Research Laboratory, RMIT University, Melbourne, Australia. }}


\footnotetext{\ddag~Present address: Department of Mechanical Engineering, University of California-San Diego, San Diego, California, USA.}

Many living materials, such as suspensions of motile microbes or of ATP-powered cytoskeletal polymers, are large ensembles of nearly identical and motile subunits that interact strongly with their neighbours. Understanding the properties of such systems  presents unique conceptual challenges. Since each elemental subunit such as a motile cell is by itself a driven--dissipative system, their collectives operate well out of equilibrium even in the absence of any external forcing.  Local fluctuations in motion further have a non-thermal origin, and the fluctuation-dissipation theorem has been shown to be inapplicable in these systems.\citep{Chen2007}

The recent development of a continuum framework that shows that these intrinsically non-equilibrium systems share universal features is therefore a significant theoretical breakthrough.\citep{Ramaswamy2010, Marchetti2013}  This theory of ``active matter" suggests that a net average local alignment of self-propelled particles must, in a continuum description, lead to a contribution to the stress tensor arising from propulsive forces or ``activity" of the particles. An interesting prediction is that particle activity must change the macroscopic viscosity of the suspension. An axisymmetric self-propelled particle exerting a net propulsive thrust along its principal axis  generates a hydrodynamic force dipole. The flow field around a single \textit{E. coli} cell has been measured to be approximately that of a positive hydrodynamic dipole \citep{Drescher2011} which forces the ambient fluid around each particle axially outward towards its two ends. The suspension viscosity for such ``pushers" is predicted to decrease  below its value for a passive suspension of inactive particles of the same size, shape and concentration. Conversely, a suspension of ``pullers" with negative hydrodynamic dipoles will have a higher viscosity than a passive suspension. These predictions\citep{Hatwalne2004} are supported by experimental observations with bacterial pushers \citep{Sokolov2009, Gachelin2013} and algal pullers.\citep{Rafai2010}  The observations appear to clearly confirm a key generic feature of the rheology of active suspensions.

Quantitative microstructural models relating particle size, shape, concentration and motility to rheological properties are just beginning to emerge,\citep{Haines2009, Saintillan2010, Saintillan2010b, Saintillan2013} but thus far there have been no systematic comparison of their predictions against experimental data.  We present here such a comparison for suspensions of wild-type strains of the microalga \textit{Dunaliella tertiolecta}, the bacterium \textit{Escherichia coli} and mouse spermatozoa. \textit{D. tertiolecta} uses its pair of flagella in a manner similar to the puller \textit{Chlamydomonas reinhardtii} studied by Raf\"{a}i \textit{et al.},\citep{Rafai2010} whereas \textit{E. coli} and the sperm use flagellar tails to push forward, like \textit{B. subtilis} studied by Sokolov and Aranson.\citep{Sokolov2009}  \textit{E. coli} cells have multiple flagella and use run-and-tumble swimming by bundling and unbundling their flagella. Sperm were cultured under conditions to induce capacitation and hyperactive swimming.\citep{Gaffney2011}  Trajectories of cells swimming were analyzed to characterize their motility under quiescent conditions.  Measurements of the extensional viscosity of live and dead cell suspensions were obtained with a novel microfluidic rheometer developed by.\citep{Bhattacharjee2011} A recent model for the rheology of dilute suspensions of self-propelled hydrodynamic dipoles \citep{Saintillan2010, Saintillan2010b} is extended here to the non-dilute regime to relate the non-Newtonian elongational viscosity of active suspensions to particle volume fraction.  We demonstrate here that despite the wide range of particle size, shape and motility characteristics, bulk mechanical behaviour of active suspensions may be accurately characterized in terms of a small number of parameters.

\section{Experiments}

\subsection{Cell culture and suspensions}
\textit{D. tertiolecta} Butcher was collected and isolated from Port Phillip Bay, Victoria, in December 2009. Cultures were maintained at 20$^\circ$C in modified F-medium  (30 g/L aquarium salt, 250 mg/L NaNO$_3$, 18 mg/L KH$_2$PO$_4$, 9 mg/L iron(III) citrate (C$_6$H$_5$O$_7$Fe), 9 mg/L citric acid (C$_6$H$_8$O$_7$), 0.2 mg/L MnCl$_2$.4H$_2$O, 0.023 mg/L ZnSO$_4$.7H$_2$O, 0.011 mg/L CoCl$_2$.6H$_2$O, 0.005 mg/L CuSO$_4$.5H$_2$O, 0.008 mg/L Na2MoO$_4$.2H$_2$O, 0.65 $\mu$ g/L H$_2$SeO$_3$ and traces of vitamin B12, biotin and thiamine \citep{Guillard1962}). 20 mL of inoculum were added to 400 mL of F-medium and incubated in 2-L glass bottles in a laboratory growth cabinet at 20 $\circ$C ± 0.1 $^\circ$C with a 16:8 light dark cycle using white fluorescent lights with a photon flux of ~60 $\mu$mol photons/ (m$^2$ s). Cultures were bubbled with air through an aquarium air stone to provide a source of inorganic carbon (CO$_2$). After 5 days, F-medium was added to bring the total culture volume to 1.5 L. Samples for experimentation were harvested after a further 6 days  during the log-phase of growth into 50 mL polypropylene capped-tubes and centrifuged at 3500 rpm at 20 $^\circ$C for 10 minutes to collect cell pellet \citep{Guillard1962}. 

Wild-type \textit{E. coli} K 12 strain was procured from ATCC, USA (\# 10798). Standard media in the form of  Luria Bertini (LB) broth (\# L3022, Sigma Aldrich; 10 g/L tryptone, 5 g/L yeast extract, 5 g/L NaCl) and/or Luria Agar (\# L2897, Sigma Aldrich) was used for bacterial culture.	 A UV-VIS spectrophotometer (\# UV-2450, Shimadzu) was used to characterize the bacterial growth by measuring absorbance / optical density at 600nm. About 0.5 mL of sterile LB broth was put into the sterile ATCC vial containing lyophilized culture. A small amount of the suspended culture (around 0.05 mL) was inoculated on to sterile Luria agar slants, and  incubated at 37 $^\circ$C for 18-20 hours. A single colony from an agar slant was transferred to 5 mL of sterile LB broth and incubated at 37 $^\circ$C for 6-7 hours with vigorous shaking (at 170 rpm), till  absorbance at 600 nm reached 0.4. About 0.3 mL of 50 wt$\%$ glycerol-water solution was added to 0.7 mL of this mid-log phase culture and stored at -73 $^\circ$C for future use. From the glycerol-freeze stock, a tiny amount is scraped off and inoculated under aseptic conditions to 5 mL of sterile LB medium. The culture was incubated at 37 $^\circ$C for 16-18 hours with vigorous shaking (at 170 rpm). A small amount (around 0.05 mL) was transferred into sterile LB media of 160 mL volume in a shake-flask. Cultures were incubated at 37 $^\circ$C for 6-7 hours with vigorous shaking (at 170 rpm).

Sperm from C57BL (wild-type) mice, extracted from cauda epididymii using the back-flushing method \citep{Gibbs2011}  was added to 5 mL pre-warmed MT6 medium (125 mM NaCl, 2.7 mM KCl, 1 mM MgCl$_2$.6H2O, 0.35 mM NaH$_2$PO$_4$.2H$_2$O, 5.5 mM glucose, 25 mM NaHCO$_3$, 1.7 mM CaCl$_2$.2H$_2$O, 60 $\mu$M bovine-serum albumin) containing methylcellulose and incubated at 37$^\circ$C for 90 min. 

Algal cultures were centrifuged at 3500 rpm for 8 min. at 20$^\circ$C and pellets were re-suspended in growth medium at desired concentrations.  Lugol's iodine (100 g/L KI, 50 g/L iodine crystals) was added to  Eppendorf tubes to kill algal cells. These were centrifuged, the supernatant removed and re-suspended in growth medium again. \textit{E. coli} cultures after log-phase growth were centrifuged at 4550 rpm for 10 min. at 4$^\circ$C, and pellets were re-suspended after weighing in a (pH 8.2) buffer of 10 mM K$_2$HPO$_4$, 0.1 mM EDTA and 0.2 wt.$\%$ glucose  to prepare suspensions of various cell volume fractions. Suspensions were exposed to UV light for 30-60 mins. to kill cells without significant lysis.  Fresh sperm samples prepared as above were first tested in capillary-thinning experiments. Standing sperm suspensions for 90 min. inactivated motility. 


\subsection{Particle tracking} 

Image-analysis was used to determine the average fractional area covered by cells at the focal plane; this was assumed to be equal to the volume fraction. In bacteria and algae, flagellar filaments were not resolved in the image analysis.  Sample droplets were placed on  teflon-coated glass slides, with a  coverslip on top. The gap width between slide and coverslip was typically 1 mm. Microscope videos for bacteria and algae were captured with 20X and  100X (Olympus) lenses, respectively, and a high-speed camera (SA5, Photron; 50 fps; 768 $\times$ 816 pixels). Swimming speeds and diffusivities  were determined by processing images (ImageJ) and cell tracking (Imaris). Sperm suspensions were loaded onto both chambers on a 2X-Cel 80 $\mu$m slides chambers covered with 2X-Cel Cover Glass (Hamilton Thorne Research). Slides were inserted into a Hamilton Throne IVOS for computer-aided semen analysis (CASA). At least 1000 sperm were counted in each chamber. Sperm motility characteristics were analyzed through image analysis as done for the algae and bacteria.

\subsection{Acoustically-Driven microfluidic extensional rheometry}  
Unlike shear rheometry, techniques for reliable measurement of fluid properties in extensional flows have been established only relatively recently.\citep{McKinley2002}  In capillary-breakup extensional rheometry (CaBER), a liquid bridge is usually first created by rapidly moving apart end-plates between which a sample drop is sandwiched. If the end-plate separation is large enough, the bridge subsequently begins to thin due to the Rayleigh-Plateau instability. The rate at which a liquid bridge thins is governed largely by the balance of the capillary stress against the inertial and viscous stresses induced by the extensional flow about the necking plane and hence it is in principle possible to extract the viscosity by monitoring the neck radius as a function of time.\citep{Szabo1997} This technique has in the past been used for highly viscous samples.\citep{McKinley2000}  Obtaining reliable measurements with low-viscosity complex fluids such as aqueous cell suspensions however presents two challenges. Firstly, mechanical motion of end-plates sets off inertial instabilities that quickly break up liquid bridges. \citep{McKinley2002} The motion within the liquid bridge  following the sudden stopping of the end-plates is complex and is not described by a simple stress balance.  Secondly, CaBER and other techniques based on capillary thinning of filaments are well established for viscoelastic fluids such as polymer solutions where liquid bridges thin exponentially in time, creating long-lived, slender and almost cylindrical filaments.\citep{Entov1997} This permits the use of a simple stress-balance equation to extract the viscoelastic fluid stress at the necking plane from just a measurement of the radius $R$ as a function of time $t$.  For fluids with little or no elasticity, directly calculating the viscosity through the stress-balance has thus far been shown to be practicable again only for very viscous fluids where a cylindrical filament can form towards the final stages of breakup.\citep{McKinley2000} Thin cylindrical filaments do form close to break-up, but imaging these require the combination of ultra-fast and very high resolution photography.\citep{Chen2002} 

\begin{figure}[ht]
\centerline{\resizebox{0.4\textwidth}{!}{\includegraphics{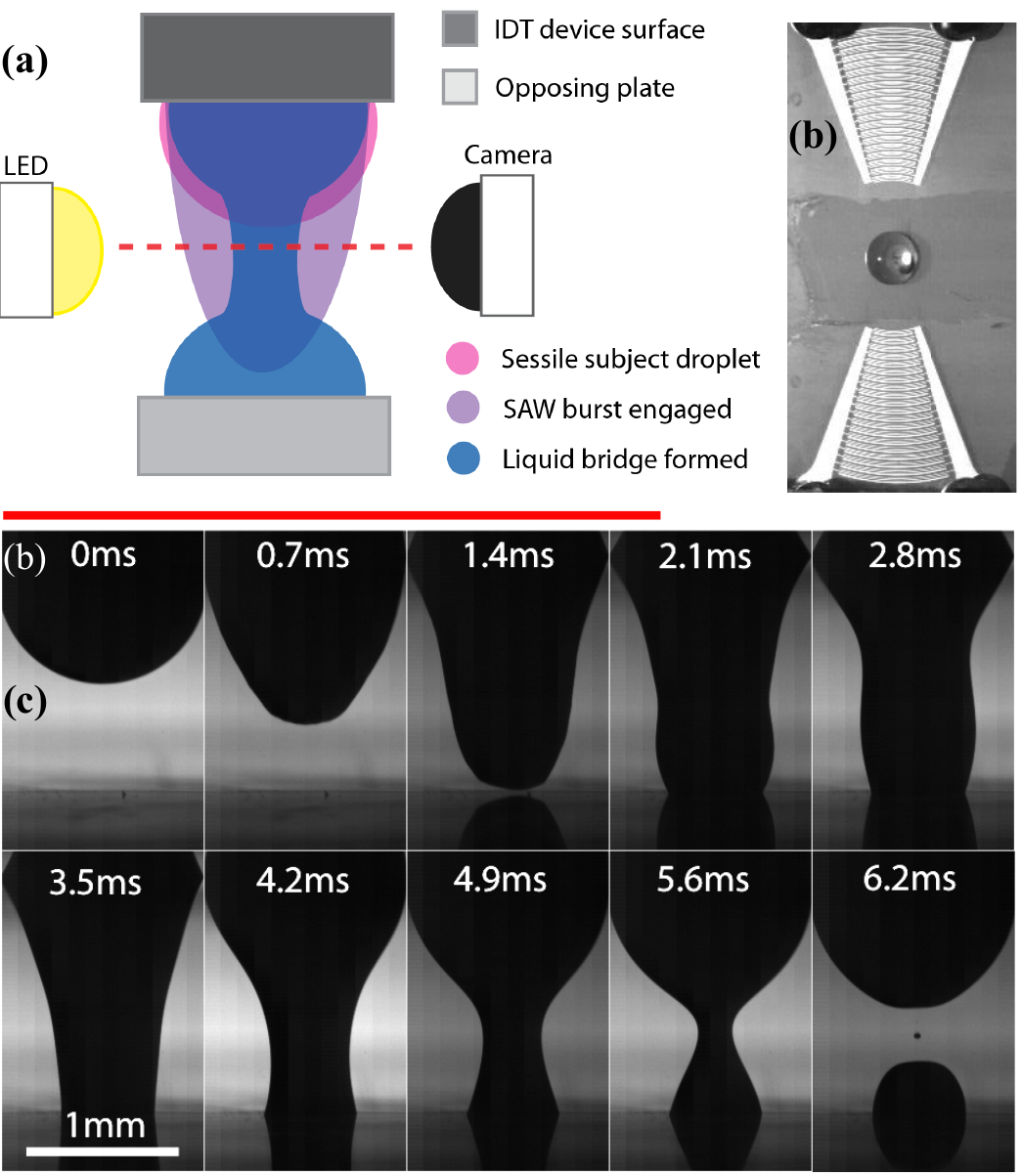}}}
\caption{\label{f:admier} (a) Schematic of experimental setup (b) Curved inter-digitated transducers for focussing SAW into a sessile droplet (c) Time-lapse images of the formation of a liquid bridge after ejection of a jet due to actuation of a sessile drop by SAW; the red-line indicates duration of the SAW pulse. }
\end{figure}

The first of these problems was overcome in a method developed by Bhattacharjee \textit{et al.},\citep{Bhattacharjee2011} wherein the liquid-bridge is created and stabilized against capillary forces initially by power input from surface acoustic radiation (Fig.~\ref{f:admier}). In our experiments, a 20 MHz waveform generator (33220A, Agilent) was used to generate a surface acoustic wave (SAW) burst triggered by a second signal generator (WF1966, NF Corporation). The latter delivered a sinusoidal signal to an RF power amplifier (411LA, ENI), providing a fixed frequency and amplitude signal near the 36.7 MHz resonance frequency of the SAW substrate. An arrangement of curved inter-digitated transducers (IDTs) bonded to a piezoelectric substrate \citep{Bhattacharjee2011} focussed Rayleigh waves towards the point where a sessile droplet (1 $\mu$l; approximately 1 mm dia.)  rests (Fig.~\ref{f:admier} (b)). Energy from the SAW leaks into the droplet causing recirculation and bulk motion, leading to an elongated liquid that bridges a gap to an opposing parallel surface located 1.5 mm away from the SAW substrate. The opposing surface was coated with teflon and was partially fouled to ensure that the jet adheres to the surface but does not spread. The SAW burst was ended after 1.5 ms which was found to be sufficient to create stable liquid bridges in all our samples. The liquid bridge then thins under the action of capillary forces, generating an extensional flow at the necking plane (Fig.~\ref{f:admier} (c)). The motion of the entire liquid bridge was captured using a high-speed camera (Photron SA5; 62000 fps; image size: 1.35 mm $\times$ 2.14 mm (192 $\times$ 304 pixels)) with a long-distance video microscope attachment (K2/SC, Infinity). The set-up is lit by a single LED lamp placed behind the filament. The radius of the neck in each image frame was obtained using standard image-analysis techniques. Initial transients were discarded in each case until the neck attained a diameter of 50 pixels (0.352 $\pm$ 0.007 mm); this was taken as the initial time ($t = 0$) for all samples.

We avoid problems with imaging thin cylindrical filaments close to break-up altogether by using neck radius data during the early stages of the thinning while the axial curvature is still large. This is furthermore advantageous when handling particle suspensions since the effect of particle interactions with the air-liquid interface on the overall dynamics at the neck is weaker when the neck radius is large compared to particle dimensions.  Rescaling the governing equations for slender but non-cylindrical Newtonian liquid-bridges \citep{Eggers1994} with the mid-filament radius $R_0$ at $t = 0$ (as defined above) and the Rayleigh time-scale,
\begin{gather}
\tauR \equiv \sqrt{ \frac{\rho \, R_0^3 }{ \gamma} }\,,
\label{e:raytime}
\end{gather}
the decay of the rescaled neck radius $R^\ast = R/R_0$ with rescaled time $t^\ast = t/\tauR$ is parameterized by the dimensionless volume and aspect ratio of the liquid bridge, and the Ohnesorge and Bond numbers
\begin{gather}
\Oh \equiv \frac{\eta}{\sqrt{\rho \, \gamma\,R_0}}\,;\quad \Bo \equiv \frac{g\,\rho \, R_0^2}{\gamma}\,,
\label{e:OhandBo}
\end{gather} 
where, $\rho$, $\gamma$ and $\eta$ are the density, surface tension coefficient and shear viscosity respectively of a fluid sample and $g$ is the gravitational acceleration. In our experiments, droplet volumes, $R_0$ and the separation between bridge surfaces were kept constant. The Ohnesorge number for our samples varied by almost three orders of magnitude, whereas the variation in $\Bo$ was relatively much smaller (around 33 \%). Therefore, the Bond number can be assumed to be relatively relatively constant. Under such conditions, the rescaled time $t^\ast_{1/2}$ taken for the filament to neck to half its initial radius (\textit{i.e.} the time to $R^\ast = 1/2$) is predominantly governed by $\Oh$. 

\begin{figure}[ht]
\centerline{\resizebox{!}{!}{\includegraphics{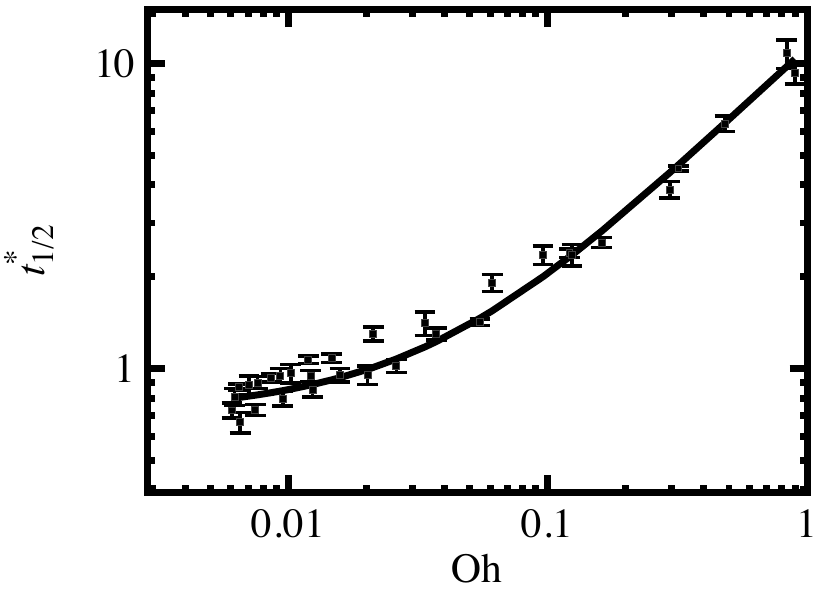}}}
\caption{\label{f:oh-calib} Variation of liquid-bridge half-times with $\Oh$: the curve is a linear least-squares fit of  Eqn.~\eqref{e:calib} through the data.}
\end{figure}

Figure~\ref{f:oh-calib} shows half-time data obtained for a range of glycerol-water and sucrose-water mixtures whose properties ($\etas$, $\gamma$ and $\rho$) were independently measured. Surface tension of suspensions were measured using a tensiometer (Analite; Selby Scientific). Shear viscosities were measured with a Haake Mars (Thermo Scientific) shear rheometer. The form of the regression curve through the data in Fig.~\ref{f:oh-calib} is inspired by the analytical solution to the inertio-viscous capillary stress balance for a cylindrical fluid filament:\citep{Entov1997,McKinley2002,Rodd2005,Tirtaatmadja2006} 
\begin{gather}
\frac{1}{2} \, \rho \dot{R}^2 = (2X - 1)\,\frac{\gamma}{R} + \frac{6 \, \etas\,\dot{R}}{R} \,,
\end{gather}
where $X$ is the ratio of the instantaneous axial tension in the liquid bridge and $2 \pi \gamma R$. If $X$ is a constant, the algebro-differential equation above can be treated as a quadratic equation above for $\dot{R}$ at any $R$ and $t$. The negative root of the rescaled equation,
\begin{gather}
\frac{d R^\ast}{d t^\ast} = - \frac{6 \,\Oh}{R^\ast} \,\left( \,\sqrt{ 1 + C\, R^\ast} - 1 \right) \,,
\end{gather}
where $C = (16 X - 8)/144$ can be integrated with $R^\ast = 1$ as the initial condition to obtain
\begin{gather}
t^\ast_{1/2} = \frac{1}{36\, C^2} \left[ \,4\,\left\{ \,\left(\,\Oh^2 + C\,\right)^{3/2} - \left(\,\Oh^2 + \frac{C}{2} \,\right)^{3/2} \right\} + 3 \, C \, \Oh \right] \,.
\label{e:thalfanal}
\end{gather}

Strictly speaking, $X$ itself depends on time; it has been shown however that, if the neck diameter is significantly smaller than the two drops at the end plates, the axial filament profile and its dynamics can be well approximated as being self-similar. As mentioned earlier, we begin taking measurements of $R$ only after the neck is well formed with a radius about half the drop radius at the end-plate. Under such conditions, a constant value of $X$ has been shown to lead to accurate predictions.  The values of $X$ given by similarity solutions under different conditions have been summarized by McKinley and Tripathy\citep{McKinley2000}. When inertia is important and $\Oh \ll 1$, $X = 0.5912$ is the value most likely to be observed in an experiment\citep{Eggers1993, Eggers1997, McKinley2000} and hence $t^\ast_{1/2} \rightarrow 0.7135$ as $\Oh \rightarrow 0$. When viscosity is dominant on the other hand, $X = 0.7127$\citep{Papageorgiou1995,McKinley2000} and $t^\ast_{1/2} \rightarrow 7.0522 \, \Oh$ as $\Oh \rightarrow \infty$. We therefore fit the experimental data obtained for intermediate $\Oh$ values with the following rational function:
\begin{gather}
t^\ast_{1/2} = \frac{0.7135 + K\, \Oh + 7.0522 \, \Oh^2}{1 + \Oh} \,,
\label{e:calib}
\end{gather}
and with $K = 14.7 \pm 0.2$ obtained by linear regression. The inverse of the function above is then used as a calibration curve to convert observed values of $t^\ast_{1/2}$ into $\Oh$ and then further into $\eeta = 3 \eta$ with the definition of the Ohnesorge number. The good agreement with experimental data of this curve derived from the analytical solution indicates that the effect of any residual flows generated by the initial SAW irradiation is negligible.


\section{Modeling}
\subsection{Rheology}
The concentration dependence of viscosity of passive suspensions below the isotropic-nematic transition is often modeled on the basis of the mean-field assumption that when a new particle is added into the free-volume of a suspension, it ``sees" the surrounding suspension as a homogeneous fluid of higher viscosity than the original suspending medium\citep{Krieger1959, Ball1980}. The incremental influence of this particle on the overall viscosity is assumed to be independent of the suspension density itself, and is therefore, in extensional flows, equal to the intrinsic viscosity coefficient of a dilute suspension, $\ieeta = \lim_{\phi \rightarrow 0}\,(\eeta - 3\,\etas)/\, (3\,\etas \,\phi)$, where $\eeta$ is the extensional viscosity of the suspension, $\etas$ is the shear viscosity of the suspending medium, and $\phi = n \, v_p$ is the volume fraction of particles,  each of volume $v_p$ and at a number density $n$. This argument leads to the Krieger-Dougherty equation  (KDE) \citep{Krieger1959, Larson1999} for the relative extensional viscosity of a non-dilute suspension,
\begin{gather}
\eetarel = \frac{\eeta}{3 \,\etas} =  \,\left( 1- \frac{\phi}{\phi_m} \right)^{-\ieeta \phi_m} \,,
\label{e:KDE}
\end{gather}
where $\phi_m$ is the maximum volume fraction at which the steady-state viscosity diverges and beyond which suspensions behave as slurries with significant solid-like characteristics.  

Recent studies by Haines \textit{et al.}\citep{Haines2009} and Saintillan\citep{Saintillan2010} have modeled the statistics of an axisymmetric self-propelled particle that stochastically changes its swimming direction while moving in an externally imposed homogeneous extensional flow, and then obtained an expression relating $\ieeta$ to motility characteristics and strain-rate.  We summarize here Saintillan's analysis for a slender rod to explain the physical significance of the key parameters in the expression for $\ieeta$ in an active suspension.  The particle contribution to the fluid stress tensor is proportional to the orientational average of its total hydrodynamic dipole which is the sum of dipoles arising from propulsion, the externally-imposed flow, and thermal fluctuations. Besides translating with a mean speed $U$, living swimmers also autonomously change direction in an apparently random manner which is modeled through a \textit{non-Brownian} rotational diffusivity, $D_r$. Passive particles (\textit{e.g.} dead swimmers) on the other hand only have a diffusivity $D_{r,\,0}$ that is solely due to thermal fluctuations. The suspending medium in turn experiences an equal and opposite reaction from the particle. In a dilute suspension undergoing a homogenous flow, the particle contribution   $\vec{\uptau}_p$ to the fluid stress tensor thus arises from an average of hydrodynamic dipole moments induced at the particle location by these forces \textit{i.e.}
\begin{gather}
\vec{\uptau}_p = n \, \left[ \langle \vec{S}^f \rangle +  \langle \vec{S}^B \rangle +  \langle \vec{S}^p \rangle \right] \,,
\end{gather}
where angular brackets represent an ensemble average over the distribution of particle orientations $\vec{p}$ and $n$ is the particle number density. In the Stokesian regime, the flow-induced dipole is proportional to the rate-of-strain tensor $\vec{E}$, with a shape-dependent proportionality constant $A$. The propulsive dipole strength is represented by another shape-dependent constant $\sigma$. The particle stress is thus derived to be
\begin{multline}
\vec{\uptau}_p = n \,A \, \left( \langle \vec{p} \vec{p} \vec{p} \vec{p} \rangle - \frac{1}{3} \, \langle \vec{p} \vec{p} \rangle \ut \right) \colon \vec{E} \\+ (3 \, n \, \kBT\, + n\,\sigma ) \, \left(\langle \vec{p} \vec{p} \rangle - \frac{1}{3} \, \ut \right) \,,
\end{multline}
where $\vec{E}$ and $\ut$ are the rate-of-strain and identity tensors, respectively, and $\kB$ is the Boltzmann constant and $T$ is the absolute thermodynamic temperature of the suspension. The orientational distribution of a slender rod is governed by a balance between frictional interactions with the medium that rotate the particle and the stochastic kicks that tend to make the  distribution isotropic. The resulting Fokker-Planck equation has a closed form analytical solution for an extensional flow at steady-state. The averages in the equation above can hence be obtained for a uniaxial extensional flow of strain-rate $\edot$. Defining the particle contribution to the suspension extensional viscosity $\eetap = (\tau_{p,\,zz} - \tau_{p,\,rr})/\edot$, where the principal stretching direction is along the $z$-axis while $r$ is the transverse radial coordinate, the specific extensional viscosity ratio is given by
\begin{multline}
\frac{\eetap}{3 \, \etas} = \frac{n \, A}{2 \,\etas} \left[ \frac{1}{2}\,\left(M + \frac{1}{3}\right) \right. \\ \left.+ \frac{3\, D_r}{\edot} \, \left( \frac{\kBT}{A D_r} + \frac{\sigma}{3 A D_r} - \frac{1}{\dB} \right)\, \left( M - \frac{1}{3}\right) \right] \,,
\label{e:intvisc1}
\end{multline}
where $\dB$ is a shape-dependent constant that arises in the coupling of the rotational rate of the particle with the straining motion of the fluid around it.\citep{HinchLeal76} The function $M$ depends on $\dB\,\edot$ and is discussed shortly. We first identify the key dimensionless model parameters and relate them to parameters that are physically measurable. 

For slender rods of hydrodynamic length $L_h$ and diameter $d$ it is known that\citep{Batchelor1970} 
\begin{gather}
A = \frac{\pi \, \etas\, L_h^3}{6\, \ln (2 \, L_h/d)}\,.
\end{gather}
In order to apply the analysis above to suspensions of microbes, we must assume that the frictional characteristics of more complex axisymmetric shapes can be represented by an effective cylinder. The total length $L$  for a flagellar swimmer is the sum of the lengths of the head and the tail when completely stretched out. A living swimmer does not of course take a fully-stretched shape, and its effective hydrodynamic length $L_h$ while swimming will in general be less than $L$.   Defining the ratio $\lambda = L_h/ L$, the dimensionless equivalent of $A$ is obtained as
\begin{gather}
\dA \,=\, \frac{A}{2 \,\etas \,v_p} \,=\,\frac{\,\pi\, \lambda^3\, L^3}{12 \, v_p\,\ln \,(2 \,\lambda \, L/d)}\,.
\end{gather}
The propulsive thrust generated by the inertialess swimmer and the corresponding frictional resistance to swimming are equal and opposite to one another, but act at different locations on the axis. A scaling argument for the strength of this propulsive dipole exerted by particle swimming at a speed $U$ suggests $\sigma \sim \etas \,L^2\,U$. Defining a geometry-dependent pre-factor $\alpha$ such that $\sigma = (\pi/2) \, \alpha\, \etas \,L^2\,U$, we obtain a dimensionless equivalent of $\sigma$ as
\begin{gather}
\dS = \frac{\sigma}{3\,A \,D_r} = \pm \,  \frac{\alpha \, \ln \,(2 \lambda \,L/d) \,U}{ \lambda^3\,L\,D_r}\,.
\label{e:dS}
\end{gather}
The sign is positive for pullers and negative for pushers, while  for passive particles, $\dS_0 = 0$. Brownian torques due to thermal fluctuations tend to relax the orientational distribution of particles to its isotropic equilibrium state; their relative strength in an active suspension is represented by 
\begin{gather}
\dG =  \frac{\kBT}{A \, D_r} = \frac{6\, \kBT\,  \ln \,(2 \,\lambda\,L/d)}{ \pi \, \etas\, \lambda^3\, L^3\,D_r}\,
\end{gather}
The Fluctuation-Dissipation Theorem (FDT) is valid for passive particles, in which case the rotational diffusivity for slender rods is 
\begin{gather}
D_{r,\,0} = \frac{\kBT}{2 A} = \frac{3 \,\kBT \ln \,( 2 \,(\lambda \,L)\,/ d)}{\pi \,\etas \,(\lambda \,L)^3} \,.
\label{e:Dr0}
\end{gather}
Therefore, $\dG_0 = 2$ for dead cells; no such simple relation is currently available for $\dG$ in active systems.

The  P\'{e}clet number --- $\Pe \equiv \edot / D_r$ (or  $\edot / D_{r,\,0}$)--- quantifies the relative ability of the extensional flow of strain-rate $\edot$ to reorient a particle against random switches in swimming direction. Using the definitions of the dimensionless parameters in Eq.~\eqref{e:intvisc1}, the intrinsic extensional viscosity is obtained as
\begin{multline}
\ieeta \,=  \lim_{\phi \rightarrow 0}\,\frac{\eetap}{3 \, \etas\, \phi}\, = \,\dA \,\left[\,\frac{1}{2}\,\left(\,M + \frac{1}{3} \,\right)\right. \\ \left. \,+\, \frac{3}{\Pe}\, \left(\,\dG \,+\, \dS \,-\, \frac{1}{\dB} \,\right)\, \left(\,M - \frac{1}{3} \,\right)\, \right] \,.
\label{e:saintillan}
\end{multline}
The function $M(x)$ is weakly monotonic in $x$ and has the form
\begin{gather}
M(x) = \frac{1}{2\,x\, D(x)} - \frac{1}{2 \,x^2}\,,
\end{gather}
where $D(x) = \exp (-x^2) \, {\int_0^x} \, {\exp (y^2)} \, dy$ is Dawson's integral. 
In Eqn.~\eqref{e:saintillan} above, $x = \sqrt{3\,\dB\,\Pe / 4 }$. As $\Pe \rightarrow 0$ therefore, $M  = 1/3 + 4/45 \, x^2 + O(x^4) = 1/3 + \dB\, \Pe/15  + O(\Pe^2) $, and when $\Pe \rightarrow \infty$, $M(x) = 1 - x^{-2} + O(x^{-4}) = 1 - 4/ (3 \,\dB\, \Pe) + O(\Pe^{-2})$.

The model above for $\ieeta$ requires, besides the solvent conditions $\kBT$ and $\etas$, independent measurement of the following parameters: the average geometric characteristics of the swimmers, $L$, $d$, $v_p$; their motility characteristics $U$ and $D_r$ (or $D_{r,\,0}$ for dead cells); and the hydrodynamic ratios $\lambda$ and  $\alpha$. The constant $\dB$ is taken to be unity, which is appropriate for slender particles of large aspect ratio.\citep{HinchLeal76,  Saintillan2010} Out of these, the solvent parameters are obtained by standard techniques and the particle size characteristics $L$, $d$ and $v_p$ are obtained through microscopy. As noted above, the value of $D_{r,\,0}$ is effectively set by invoking the FDT and using $\dG_0 = 2$. We discuss below the estimation of the motility parameters, $U$ and $D_r$. It is more difficult to directly determine the ratios $\lambda$ and $\alpha$, which are therefore treated as free parameters to be obtained by comparing model predictions to experimental data for $\eetarel$. Keeping in mind their physical significance, we assess later if the values for  $\lambda$ and $\alpha$ obtained thus are plausible.

\subsection{Active diffusivity}
In the absence of flow, the trajectory of an active particle swimming in with a speed $U$ along its principal axis with random reorientations of that axis can be modeled by a pair of Langevin equations for its instantaneous  position $\vec{r}(t)$ and orientation.
The rescaled mean-squared displacement for such a particle in two dimensions is\citep{tenHagen2011}
\begin{gather}
\widetilde{\mathrm{MSD}} =   \frac{\langle\, (\vec{r}(t) - \vec{r}(0))^2\,\rangle}{ (U/D_r)^2} =  \,4 \,\xi \,\tilde{t} \,+ \,2  \,\left( \tilde{t} - 1 + e^{- \tilde{t}} \right) \,,
\label{e:MSDtheory}
\end{gather}
where $\tilde{t} = D_r\,t$ and $\xi = D_t \, D_r/U^2$. The rescaled translational diffusivity $\xi$ controls the shape of the $\widetilde{\mathrm{MSD}}$-versus-$\tilde{t}$ curves:
\begin{align}
\widetilde{\mathrm{MSD}} \approx 
\begin{cases}  
\,(4 \,\xi \, + \tilde{t}) \,\tilde{t} \,, \text{ if } \tilde{t} \ll 1\\
(4 \,\xi \, + 2)\, \tilde{t} \,, \text{ if } \tilde{t} \gg 1\,.
\end{cases}
\end{align}
If translational diffusion is dominant and $\xi \gg 1$, then $\widetilde{\mathrm{MSD}} \approx 4 \,\xi\, \tilde{t}$ always. When $0 < \xi < 1$ on the other hand, one observes diffusive-ballistic-diffusive behaviour. At short times when $ \tilde{t} \ll 4 \,\xi$, $\widetilde{\mathrm{MSD}}  \approx 4 \,\xi\, \tilde{t}$. At long time scales too, the behaviour is diffusive and $\widetilde{\mathrm{MSD}}  \approx (4 \,\xi + 2)\,\tilde{t}$. Therefore, if the effective long-time translational diffusivity $D_\mathrm{eff}$ is defined such that $\mathrm{MSD} = 4\,D_\mathrm{eff}\,t$, then $D_\mathrm{eff} \,= \,D_t \,+\, (U^2/ 2\,D_r)$. When $4 \,\xi \ll  \tilde{t} \lesssim 1$, the motion is ballistic with $\widetilde{\mathrm{MSD}} \approx \tilde{t}^2$. Values of $D_r$, $D_t$ and $U$ can be obtained from shift-factors by shifting experimental $\mathrm{MSD}$-versus-$t$ data to match one of the dimensionless theoretical curves on a log-log plot.

\section{Results and discussion}
\begin{figure}[t]
\centerline{\resizebox{8.3cm}{!}{\includegraphics{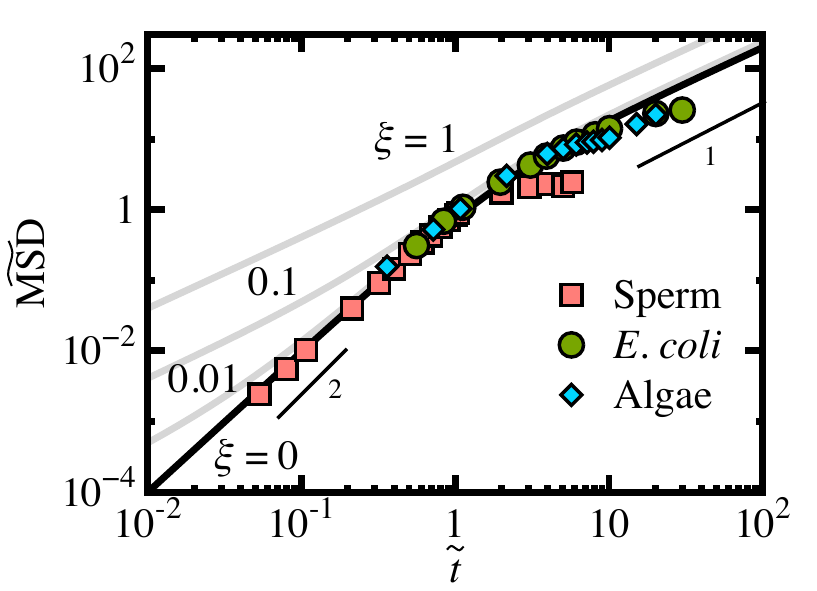}}}
\caption{\label{f:MSD} Comparison of experimental data for the mean-squared displacement for algae, \textit{E. coli} and mouse sperm with predictions for various values of the relative translational diffusivity $\xi$;  $\widetilde{\mathrm{MSD}} = \mathrm{MSD}\, (D_r/ U)^2$ and $\tilde{t} = D_r\, t$.}
\end{figure}

Experimental MSD data was obtained using suspensions of very low volume fractions with cell number densities smaller than $L^{-3}$. We find that this data for all the three organisms resembled the $\xi=0$ curve, indicating that translational diffusivity during swimming has a negligible influence (Fig.~\ref{f:MSD}). The values of swimming speed and diffusivity thus extracted are reported in Table~\ref{t:diffparams}. The value of $D_r$ for algae is comparable to those observed by Rafa\"{i} \textit{et. al}\citep{Rafai2010} in another blue-green algal species \textit{C. reinhardtii}: they measured an average speed of  $U = 40$ $\mu$m/s, and a long-time effective diffusivity of $D_\mathrm{eff} \approx 995$ $\mu$m$^2$/s, which corresponds to a rotational diffusivity of $D_r \approx 0.8$ s$^{-1}$ assuming that the true translational diffusivity is negligible in that species as well. For \textit{E. coli}, analysis of the 3D cell-tracking data suggests  $D_r \approx 3.5$ s$^{-1}$\citep{Saragosti2012}, although a much lower value of $D_r = 0.057$ s$^{-1}$ was reported by Drescher \textit{et al.}\citep{Drescher2011} Some of this variability can be attributed to differences in strains and the media and protocols used for cell-tracking. In addition, there are also significant differences in the analyses of the run-and-tumble motion of \textit{E. coli} cells to assign effective diffusivities. Although computer-aided analysis of sperm motility is well established in mammalian reproductive biology,\citep{Mortimer1997} to our best knowledge, mean-square displacements have not been analyzed to determine effective diffusivities.   The values of $U$ extracted by fitting the theoretical prediction in Eqn.~\eqref{e:MSDtheory} through the mean-squared displacement data for the algae and bacteria are close to the values measured by direct observation (45 and 4.7 $\mu$m/s, respectively). In the case of sperm however, the extracted value is significantly different from the direct measurement of an average speed 70 $\mu$m/s for the motion of the sperm head which is comparable to  average head speeds of around 100 - 150 $\mu$m/s in hyperactive mice sperm.\citep{Qi2007, Olson2011}. This difference could be due to the fact that the sperm head oscillates about the mean trajectory of the cell.\citep{Mortimer1997}

\begin{table}[h]
\small
\caption{\label{t:diffparams} Geometric and motility characteristics}
\begin{tabular*}{0.5\textwidth}{@{\extracolsep{\fill}}lccc}
    \hline
 & \textbf{Algae} & \textbf{Bacteria} & \textbf{Sperm} \\
     \hline
\multicolumn{4}{c}{}  \\[1mm]   
$L$ ($\mu$m) & 36 & 13 & 95 \\
$d$ ($\mu$m) & 2.8 & 1  &  3   \\
$L/d$ ($\mu$m) & 12.9 & 13  &  32   \\
$U$ ($\mu$m/s) & 30 & 5.5  &  31   \\
$D_r$ (1/s) &  3 & 10  &  0.8   \\[1mm] 
\hline
\end{tabular*}
\end{table}

\begin{figure*}[p]
\centerline{\resizebox{!}{!}{\includegraphics{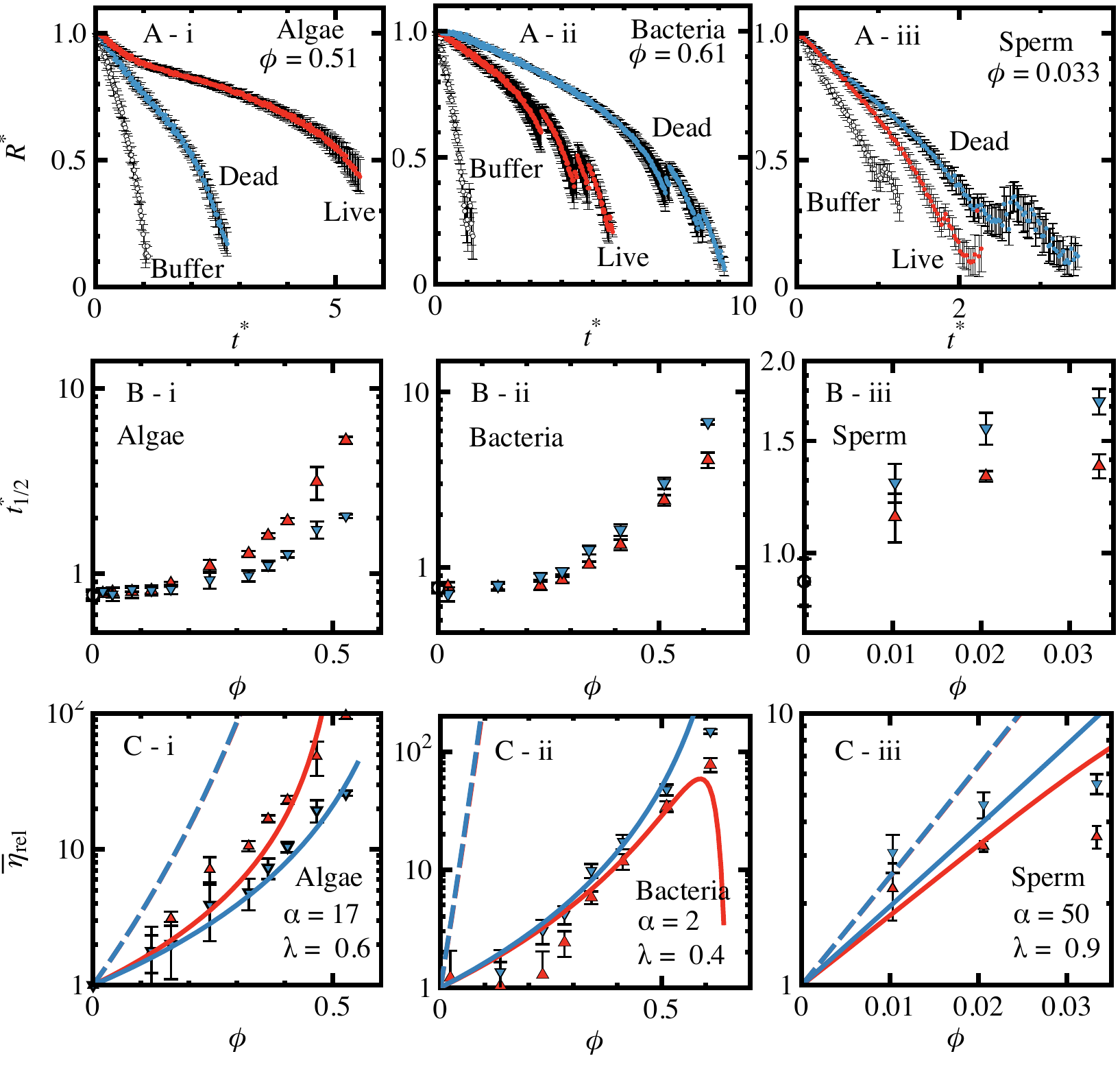}}}
\caption{\label{f:KDvsexpt} Top panel: radial decay observed during capillary thinning for the most concentrated algal, bacterial and sperm suspensions studied; middle panel: variation of dimensionless capillary thinning half-times with particle  volume fraction; bottom panel: comparison of predictions of relative extensional viscosity $\eeta_\mathrm{rel} = \eeta/ (3 \etas)$ with experimental values extracted from measured $t^\ast_{1/2} = t_{1/2}/\tauR$. In B and C, up- and down-triangles represent data for live and dead cell suspensions, respectively. Continuous curves in the bottom panel have been obtained with values of $\alpha$ and $\lambda$ as shown on the plots whereas dashed curves are predictions with both these parameters set to unity.}
\end{figure*}

Figure~\ref{f:KDvsexpt} presents the key results in our study. The top panel in the Figure shows the evolution of the neck radius during capillary thinning for the highest volume fractions studied in each species; data at lower volume fractions is presented in the Supplementary Information.$^\dag$ We observe that capillary thinning of samples of live algal suspensions progresses more slowly than of samples of dead cell suspensions at the same volume fraction (top-panel, Fig.~\ref{f:KDvsexpt}). The opposite behaviour is observed in bacterial and sperm suspensions. This is in line with the expectation that motility of pullers like \textit{D. tertiolecta} tends to increase the viscosity whereas in pushers such as \textit{E. coli} and sperm, viscosity decreases. The middle and bottom panels in Fig/~\ref{f:KDvsexpt} shows that the differences between the radius-versus-time data, and thus between the viscosities, of live and dead cell suspensions. These differences in \textit{E. coli} are small but significant relative to experimental uncertainty. In comparison,  the differences are larger in algae over a similar range of volume fractions.  Data for sperm could only be obtained at low concentrations due to the small sample volumes collected from mice and limitations in concentrating them further. However, even at these low concentrations, the effect of motility on viscosity appears clear.The difference in viscosity between live and dead cell samples appears to increase with volume fraction for all three species. 

To use the model to understand these trends, we first need to determine the strain rates in the experiments. As is well known, the instantaneous strain-rate $\edot = - 2 \, d \ln R/ \,dt$ at the necking plane in CaBER experiments cannot be directly controlled but are determined by the liquid bridge dimensions and fluid properties. The radial decay in all our samples was observed to be approximately linear when $ t < t_{1/2}$. The instantaneous strain-rate in this observation period for linear radial decay increases from $1/t_{1/2}$ to $2/t_{1/2}$. We can thus estimate the average strain-rate from the obesrved $t_{1/2}$ as $\edota \approx  \epsilon_{1/2} \,/ t_{1/2} = 1.4/t_{1/2} $ since the the Hencky strain at $t_{1/2}$ for a linear radial decay can be shown to be $\epsilon_{1/2} = 2\,\ln 2$. It is therefore clear from the systematic increase in $t_{1/2}$ with concentration (Fig.~\ref{f:KDvsexpt} B (i --- iii)) that the average strain-rate $\edota$ decreases with concentration as the increasing viscosity slows down capillary thinning.   The strain-rate at any concentration in the range covered in the experiments was obtained using a cubic polynomial fit through the experimental $\edota$-vs-$\phi$ data.$^\dag$ Values of $\edota$ in our experiments ranged from about 250 s$^{-1}$ at higher concentrations to 2000 s$^{-1}$ for pure buffers.

\begin{table}[h]
\small
\caption{\label{t:params}Model parameters}
\begin{tabular*}{0.5\textwidth}{@{\extracolsep{\fill}}lccc}
    \hline
 & \textbf{Algae} & \textbf{Bacteria} & \textbf{Sperm} \\
     \hline
\multicolumn{4}{c}{Free parameters}\\[1mm] 
$\alpha$ & 15 & 1 & 30 \\
$\lambda$ & 0.6 & 0.3 & 0.7 \\[1mm] 
$\phi_m$ & 0.7 & 0.7 & 0.7 \\[1mm]
\multicolumn{4}{c}{Calculated parameters}\\[1mm]
$\dA$ & 5.2 & 6.7 & 100 \\
$\dG$ & $4.5 \times 10^{-3}$ & $1.2 \times 10^{-2}$ & $5.4 \times 10^{-5}$ \\
$\dS$ & 40 & -2.7 & -97 \\
$F$ (pN) & 46 & 0.6 & 260 \\[1mm]
$D_{r,\,0}$ (1/s) &   $6.8 \times 10^{-3}$ & $6.1 \times 10^{-2}$  &   $2.2 \times 10^{-5}$  \\
\hline
\end{tabular*}
\end{table}

Values of model parameters and other derived quantities for the three species are listed in Table~\ref{t:params}.The ratio $\lambda$ of the hydrodynamic length to the total end-to-end length $L$ is a parameter that is estimated by visually bringing the predictions for dead cell suspensions into agreement with the experimental data. Since we are interested in order of magnitude estimates for this and the other free parameters $\alpha$ and $\phi_m$ in the model, a more rigorous least-squares fit was not pursued. The estimation of the parameter $\lambda$ is not independent of the maximum packing fraction $\phi_m$. While $\phi_m \approx 0.63$ is often used for spheres, there is no consensus on its value for anisotropic rod-like particles. It has been shown that $\phi_m$ for such systems depends on particle aspect ratio for dilute systems.\citep{Larson1999} More concentrated suspensions of rod-like particles undergo an isotropic-to-nematic transition which is expected to occur at $\phi \sim O((L/d)^{-1})$ at equilibrium.\citep{Larson1999,doiedw} The aspect ratios  (based on the total end-to-end length; Table~\ref{t:diffparams}) of all three species are quite large. The concentrations of the algal and bacterial suspensions in our experiments are thus possibly well above the isotropic-nematic transitions for those systems. Although the volume fractions of the sperm suspensions are low, significant orientational effects due to interparticle interactions cannot be ruled out due to their large aspect ratios. A value of $\phi_m = 0.7$ gives reasonable agreement of model prediction with dead cell data for all three species. With $\lambda$ and $\phi_m$ determined in this manner, $\alpha$ is estimated to match predictions for live cells with experimental data for $\eetarel$. 

\begin{figure}[t]
\centerline{\resizebox{8.3cm}{!}{\includegraphics{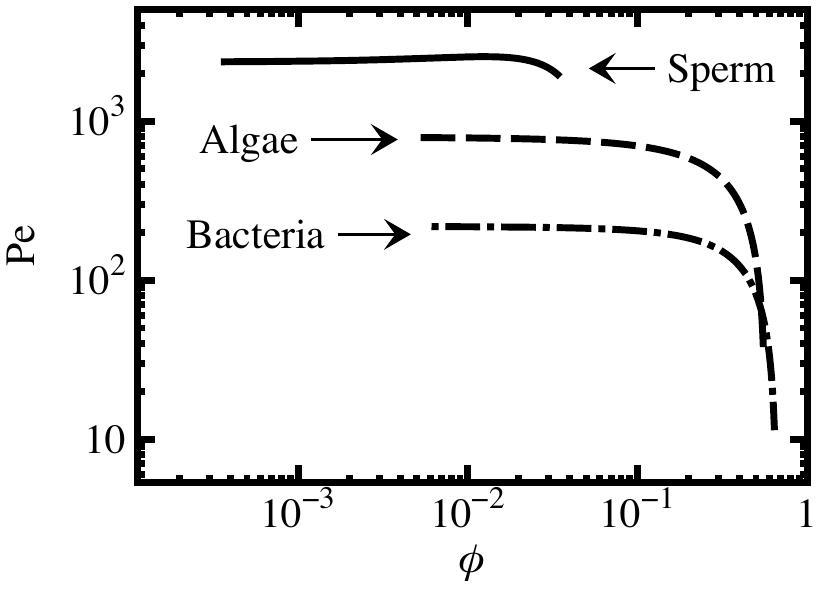}}}
\caption{\label{f:pe} Experimentally observed variation of $\Pe$ with volume fraction in live samples; the curves have been obtained by polynomial regression of strain-rate data.$^\dag$ }
\end{figure}

The predictions of $\eetarel$ in Fig.~\ref{f:KDvsexpt} were generated using the KDE after first calculating the intrinsic viscosities $\ieeta$ with Eqn.~\eqref{e:saintillan} for the ranges of P\'{e}clet numbers observed for live and dead cells.
The P\'{e}clet numbers calculated as $\Pe = \edota \, D_r$ with the rotational diffusivities for live cells in Table~\ref{t:diffparams} are  plotted in Fig.~\ref{f:pe}. The P\'{e}clet number governs the competition between flow and diffusion and the values of $\Pe \gg 1$  in Fig.~\ref{f:pe} indicate that the flow completely overcomes the tendency for diffusion to make the orientational distribution isotropic and that particles are aligned along the principal stretching axis of the unaxial extensional flow. In such an aligned state, the relative contributions of the flow-induced and propulsive dipoles to the total stress and hence the viscosity is determined by the magnitude of the parameter $\dS$ in an active suspension. In fact, in the case of dead cells for which this parameter is zero, the flow wins by default. Moreover, the much smaller rotational diffusivities (Table~\ref{t:params}) of dead cells  result in even larger P\'{e}clet numbers than in live cell suspensions at comparable strain rates. The values of $\ieeta$ for dead cells in each species are thus nearly identical to the limiting value  $\ieeta^\infty 2 \,\dA/ 3$ contributed solely by the flow-induced dipole as $\Pe \rightarrow \infty$.
In active suspensions on the other hand, even when cells are highly aligned by the flow when $\Pe \gg 1$, the contribution of propulsive forces can be relatively large and $\ieeta$ is significantly different from $\ieeta^\infty$. This is the principal reason behind the significant differences observed between live and dead cell samples in Fig.~\ref{f:KDvsexpt} despite the large strain rates and $\Pe$. Moreover, with increasing cell volume fraction, not only are there more cells per unit volume contributing propulsive stresses, the strain-rate and $\Pe$ numbers decrease (Fig.~\ref{f:pe}) boosting the relative contribution of active dipole.

\begin{figure}[t]
\centerline{\resizebox{8.3cm}{!}{\includegraphics{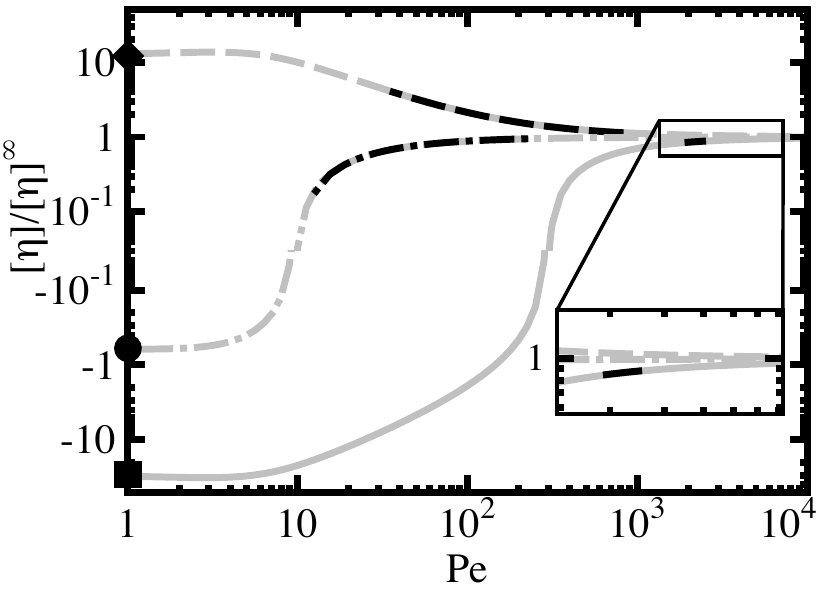}}}
\caption{\label{f:ivratio} Effect of flow strength on intrinsic viscosity of live cell suspensions of sperm (continuous), algae (dashed) and bacteria (dot-dashed curve): the dark portions of the curves represent the experimentally observed ranges of $\Pe$ values; the symbols represent the zero-$\Pe$ limit, $\ieeta^0/\ieeta^\infty$ obtained with the model parameters in Table~\ref{t:params}.}
\end{figure}

The effect of activity on $\eetarel$ in Fig.~\ref{f:KDvsexpt} appears strongest in algae because of the combination of a large propulsive dipole and particle concentrations. The dipole is much weaker in \textit{E. coli} pushers, leading to a smaller effect of activity, although they have the largest swimming diffusivity and lowest $\Pe$ values among the three species. In contrast, $\dS$ in sperm appears to be so large that their live suspensions show a clearly measurable effect of motility even at very low cell densities and despite very large $\Pe$ values. Interestingly, the negative values of $\dS$ for the pushers \textit{E. coli} and sperm are predicted to lead to negative $\ieeta$ at sufficiently low strain-rates (Fig.~\ref{f:ivratio}), which will result in live cell suspensions being less viscous than the suspending medium. Although such low strain-rates cannot be practically realized with extensional rheometers currently, they are  accessible in shear rheometers. This intriguing effect  has been demonstrated by Gachelin \textit{et al.}\citep{Gachelin2013} with \textit{E. coli} in a microfluidic device. This  effect is likely to be much more dramatic with sperm suspensions: the parameters we have obtained for sperm lead to a very large negative zero-$\Pe$ limit $\ieeta^0 = \dA\,(2/9 + \dB/15 \, ( \dG + \dS - 1/\dB) \,)= -1.9 \times 10^3$. 

How reasonable are the values (in Table~\ref{t:params} of the key free parameters: the prefactor $\alpha$ in Eqn.~\eqref{e:dS} for the active dipole, and the ratio $\lambda$ of the hydrodynamic length to the total head-tail length of a swimmers? If both these parameters are set to unity, the values of the activity $\dS$ are too low  for all three species and predictions for live and dead cell suspensions are virtually indistinguishable, and these  predictions are substantially different from the experimental data (dashed curves in Fig.~\ref{f:KDvsexpt} C (i --- iii) ).  The values of $\lambda \lesssim O(1)$ required to obtain agreement with experiment appear reasonable given that flagella are never fully stretched out in swimming cells. Hydrodynamic interactions between the head and tail cause the parameter $\alpha$ to be strongly geometry dependent. This can be seen by modeling a flagellar swimmer modeled as an asymmetric rigid dumbbell with Stokeslets of different hydrodynamic radii located at the head and tail respectively (Appendix A).\citep{Dunstan2012}  We obtain for such a dumbbell $\alpha = 12 \, (L_{ht}/L)^2\,/\, [(L_{ht}/a_h) - 3/2]$, where $L_{ht}$ is the distance between the head and tail centres and $a_h$ is the hydrodynamic radius of the head. Although finite-size corrections can be expected to modify the singular behaviour when $L_{ht}/ a_h = 3/2$, this result suggests that $\alpha$ can vary over a wide range depending on the ratio $L_{ht}$ and $a_h$. Viewed in this light, the variation in values of $\alpha$ for the three very differently shaped cells appears plausible. 

It if further possible to estimate propulsive forces (using the definition of $\dS$ in Eqn.~\eqref{e:dS} and its values in Table~\ref{t:params}) as $F = \sigma/L_h$. Bayly \textit{et al.}\citep{Bayly2011} analyzed the flagellar stroke of a single \textit{C. reinhardtii} using resistive force theory to estimate an average power dissipation of about 5 fW, which when combined with an average speed of around 40 $\mu$m/s in that species,\citep{Rafai2010} yields $F = 125$ pN. Drescher \textit{et al.}\citep{Drescher2011} estimated a propulsive force of $F = 0.43$ pN directly in a wild-type \textit{E. coli} cell by measuring the flow field around it during a straight run. Schmitz \textit{et al.}\citep{Schmitz2000} used a microprobe to measured the force required to stall the motion of a bull sperm flagellum to be $F = 250$ pN. These values from direct single cell or flagellum measurements are comparable to the values of $F$ in Table~\ref{t:params}).

\section{Conclusions}
Our experimental observations with a surface-acoustic-wave driven microfluidic rheometer indicate that particle motility has a clearly measurable influence on the rheology of suspensions. Capillary thinning of liquid bridges proceeded more slowly in suspensions of algal pullers than those of dead cells at the same volume fraction, whereas bacterial and sperm pushers tended to hasten thinning. The difference in the effective viscosity between suspensions of live and dead cells was found to systematically increase with concentration. Predictions with a model that combined the Krieger-Dougherty equation for the relative viscosity of suspensions with an equation derived by Saintillan\citep{Saintillan2010} for the intrinsic extensional viscosity of active suspensions were found to be in good qualitative agreement with the experimental observations. Our results show that the propulsive dipole even in weak swimmers such as \textit{E. coli} can contribute significantly to fluid stresses even at high P\'{e}clet numbers when flow dominates over swimming noise. While  parameters such as the properties of the suspending medium, the physical dimensions of the cells, and their swimming speeds and effective rotational diffusivities   could be determined from independent measurements, model predictions depend crucially  on hydrodynamic details such as the effective hydrodynamic length of the swimmer and the contribution of hydrodynamic interactions between the head and tail to the propulsive dipole moment. Values of these parameters required to match predictions with experimental data lead to estimates for propulsive forces in the three species that are in line with data in literature on direct force measurements with single cells or flagella. This suggests that it is possible to develop a  better models for the nonlinear rheology of microswimmer suspensions by combining more accurate alternatives for the Krieger-Dougherty formulation describing concentration effects with further refinements of the dilute solution theory\citep{Haines2009, Saintillan2010, Saintillan2010b} for the viscosity of active suspensions that more accurately reflect the hydrodynamic characteristics of swimming particles and their propulsion. Such models open up the possibility of routinely and precisely extracting propulsive forces from rheometry of active suspensions.

\section{Acknowledgments}
The authors thank Sharadwata Pan (IITB-Monash Research Academy, Mumbai, India), David Hill (University of Melbourne, Australia) and Michael K. Danquah (Curtin University, Sarawak, Malaysia) for training AGM in microbiological techniques for culturing and preparing live/dead-cell suspensions of \textit{E. coli} and \textit{D. tertiolecta}.

\appendix

\section{Self-propelled asymmetric dumbbells}
\begin{figure}[ht]
\centerline{\resizebox{8.3cm}{!}{\includegraphics{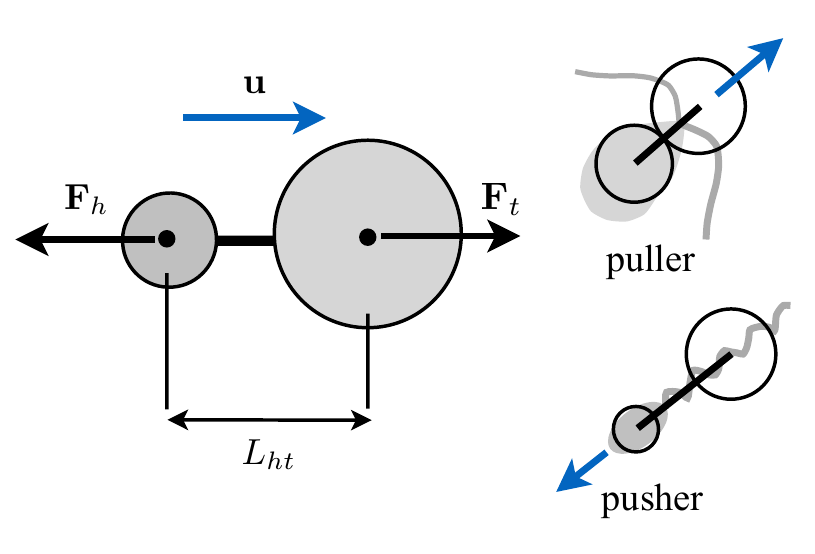}}}
\caption{\label{f:sketch} Self-propelled asymmetric dumbbell model of a microswimmer with a flagellar tail}
\end{figure}

A swimmer is modeled as a rigid asymmetric dumbbell with Stokeslets of hydrodynamic radii $a_h$ and $a_t$ located at the hydrodynamic centres of the head and tail respectively \citep{Dunstan2012}. The rigid swimmer translates with a velocity $\vec{u}$ relative to the unperturbed ambient fluid as a result of the actuation of the tail. The tail generates a thrust of $\vec{F}_p$. The total hydrodynamic force \textit{on} the tail is the sum of $\vec{F}_p$ and its frictional resistance to the translational motion:
\begin{gather}
\vec{F}_t = - \zt \,( \vec{u} + \bom \cdot \vec{F}_h ) \, + \, \vec{F}_p \,,
\end{gather}
where $\zt = 6\, \pi \,a_t\, \etas$ is the drag coefficient of the tail and  $\bom$ is the tensor describing the hydrodynamic interaction tensor between the head and tail such that the velocity perturbation at the tail due to the hydrodynamic force $\vec{F}_h$ on the head is $-\bom \cdot \vec{F}_h$ \citep{dpl2}. The head only experiences frictional resistance, and
\begin{gather}
\vec{F}_h = - \zh \,( \vec{u} + \bom \cdot \vec{F}_t )  \,.
\end{gather}
The total hydrodynamic force on the inertialess swimmer must be zero in the absence of external forces:
\begin{gather}
\vec{F}_h + \vec{F}_t = \vec{0} \,.
\end{gather}
The three equations above can be solved for $\vec{F}_h$, $\vec{F}_t$ and $\vec{F}_p$, given the frictional characteristics ($\zh$, $\zt$ and $\bom$) and the velocity $\vec{u}$. Assuming that all forces and the velocity are directed along the axis of the dumbbell, if $F_h$, $F_t$ and $F_p$, and $u$ are the axial force and velocity components, we obtain,
\begin{gather}
F_t = - F_h = \frac{\zh}{1 - \Omega \, \zh} \, u \,, \\
\end{gather}
where $\Omega = 1/ (4 \,\pi \,\etas \,L_p)$ for Oseen-Burgers hydrodynamic interaction, and $L_p$ is the distance between the head and tail centres. The magnitude of propulsive dipole is then
\begin{gather}
| \sigma | \,=\, |F_h| \,L_p \,=\, \frac{ 12 \,(\pi/2) \,\etas \,L_p^2 \,|u|}{(L_p/\,a_h) - (3/2)} \,.
\end{gather}
With  $\alpha$ defined as before \textit{i.e.}  $| \sigma | = (\pi/2)\, \alpha \,\etas \,L^2\, |u|$ where $L$ is the total end-to-end length of the swimmer, we obtain
\begin{gather}
\alpha = \frac{12 \,(L_p/L)^2}{(L_p/ a_h) - (3/2)} \,.
\end{gather}

\footnotesize{
\bibliography{bib} 

\providecommand*{\mcitethebibliography}{\thebibliography}
\csname @ifundefined\endcsname{endmcitethebibliography}
{\let\endmcitethebibliography\endthebibliography}{}
\begin{mcitethebibliography}{42}
\providecommand*{\natexlab}[1]{#1}
\providecommand*{\mciteSetBstSublistMode}[1]{}
\providecommand*{\mciteSetBstMaxWidthForm}[2]{}
\providecommand*{\mciteBstWouldAddEndPuncttrue}
  {\def\EndOfBibitem{\unskip.}}
\providecommand*{\mciteBstWouldAddEndPunctfalse}
  {\let\EndOfBibitem\relax}
\providecommand*{\mciteSetBstMidEndSepPunct}[3]{}
\providecommand*{\mciteSetBstSublistLabelBeginEnd}[3]{}
\providecommand*{\EndOfBibitem}{}
\mciteSetBstSublistMode{f}
\mciteSetBstMaxWidthForm{subitem}
{(\emph{\alph{mcitesubitemcount}})}
\mciteSetBstSublistLabelBeginEnd{\mcitemaxwidthsubitemform\space}
{\relax}{\relax}

\bibitem[Chen \emph{et~al.}(2007)Chen, Lau, Hough, Islam, Goulian, Lubensky,
  and Yodh]{Chen2007}
D.~Chen, A.~Lau, L.~Hough, M.~Islam, M.~Goulian, T.~Lubensky and A.~Yodh,
  \emph{Phys. Rev. Lett}, 2007, \textbf{99}, 148302\relax
\mciteBstWouldAddEndPuncttrue
\mciteSetBstMidEndSepPunct{\mcitedefaultmidpunct}
{\mcitedefaultendpunct}{\mcitedefaultseppunct}\relax
\EndOfBibitem
\bibitem[Ramaswamy(2010)]{Ramaswamy2010}
S.~Ramaswamy, \emph{Ann. Rev. Cond. Mat. Phys.}, 2010, \textbf{1},
  323--345\relax
\mciteBstWouldAddEndPuncttrue
\mciteSetBstMidEndSepPunct{\mcitedefaultmidpunct}
{\mcitedefaultendpunct}{\mcitedefaultseppunct}\relax
\EndOfBibitem
\bibitem[Marchetti \emph{et~al.}(2013)Marchetti, Joanny, Ramaswamy, Liverpool,
  Prost, Rao, and Simha]{Marchetti2013}
M.~C. Marchetti, J.~Joanny, S.~Ramaswamy, T.~B. Liverpool, J.~Prost, M.~Rao and
  R.~A. Simha, \emph{Rev. Mod. Phys.}, 2013, \textbf{85}, 1143--1189\relax
\mciteBstWouldAddEndPuncttrue
\mciteSetBstMidEndSepPunct{\mcitedefaultmidpunct}
{\mcitedefaultendpunct}{\mcitedefaultseppunct}\relax
\EndOfBibitem
\bibitem[Drescher \emph{et~al.}(2011)Drescher, Dunkel, Cisneros, Ganguly, and
  Goldstein]{Drescher2011}
K.~Drescher, J.~Dunkel, L.~H. Cisneros, S.~Ganguly and R.~E. Goldstein,
  \emph{Proc. Natl. Acad. Sci. USA}, 2011, \textbf{108}, 10940--10945\relax
\mciteBstWouldAddEndPuncttrue
\mciteSetBstMidEndSepPunct{\mcitedefaultmidpunct}
{\mcitedefaultendpunct}{\mcitedefaultseppunct}\relax
\EndOfBibitem
\bibitem[Hatwalne \emph{et~al.}(2004)Hatwalne, Ramaswamy, Rao, and
  Simha]{Hatwalne2004}
Y.~Hatwalne, S.~Ramaswamy, M.~Rao and R.~Simha, \emph{{Rheology of
  active-particle suspensions}}, 2004\relax
\mciteBstWouldAddEndPuncttrue
\mciteSetBstMidEndSepPunct{\mcitedefaultmidpunct}
{\mcitedefaultendpunct}{\mcitedefaultseppunct}\relax
\EndOfBibitem
\bibitem[Sokolov and Aranson(2009)]{Sokolov2009}
A.~Sokolov and I.~Aranson, \emph{Phys. Rev. Lett.}, 2009, \textbf{103},
  148101\relax
\mciteBstWouldAddEndPuncttrue
\mciteSetBstMidEndSepPunct{\mcitedefaultmidpunct}
{\mcitedefaultendpunct}{\mcitedefaultseppunct}\relax
\EndOfBibitem
\bibitem[Gachelin \emph{et~al.}(2013)Gachelin, Mi{\~n}o, Berthet, Lindner,
  Rousselet, and Clement]{Gachelin2013}
J.~Gachelin, G.~Mi{\~n}o, H.~Berthet, A.~Lindner, A.~Rousselet and E.~Clement,
  \emph{Phys. Rev. Lett.}, 2013, \textbf{110}, 268103\relax
\mciteBstWouldAddEndPuncttrue
\mciteSetBstMidEndSepPunct{\mcitedefaultmidpunct}
{\mcitedefaultendpunct}{\mcitedefaultseppunct}\relax
\EndOfBibitem
\bibitem[Rafa{\"{i}} \emph{et~al.}(2010)Rafa{\"{i}}, Jibuti, and
  Peyla]{Rafai2010}
S.~Rafa{\"{i}}, L.~Jibuti and P.~Peyla, \emph{Phys. Rev. Lett.}, 2010,
  \textbf{104}, 098102\relax
\mciteBstWouldAddEndPuncttrue
\mciteSetBstMidEndSepPunct{\mcitedefaultmidpunct}
{\mcitedefaultendpunct}{\mcitedefaultseppunct}\relax
\EndOfBibitem
\bibitem[Haines \emph{et~al.}(2009)Haines, Sokolov, Aranson, Berlyand, and
  Karpeev]{Haines2009}
B.~Haines, A.~Sokolov, I.~Aranson, L.~Berlyand and D.~Karpeev, \emph{Phys. Rev.
  E}, 2009, \textbf{80}, 041922\relax
\mciteBstWouldAddEndPuncttrue
\mciteSetBstMidEndSepPunct{\mcitedefaultmidpunct}
{\mcitedefaultendpunct}{\mcitedefaultseppunct}\relax
\EndOfBibitem
\bibitem[Saintillan(2010)]{Saintillan2010}
D.~Saintillan, \emph{Phys. Rev. E}, 2010, \textbf{81}, 056307\relax
\mciteBstWouldAddEndPuncttrue
\mciteSetBstMidEndSepPunct{\mcitedefaultmidpunct}
{\mcitedefaultendpunct}{\mcitedefaultseppunct}\relax
\EndOfBibitem
\bibitem[Saintillan(2010)]{Saintillan2010b}
D.~Saintillan, \emph{Exp. Mech.}, 2010, \textbf{50}, 1275--1281\relax
\mciteBstWouldAddEndPuncttrue
\mciteSetBstMidEndSepPunct{\mcitedefaultmidpunct}
{\mcitedefaultendpunct}{\mcitedefaultseppunct}\relax
\EndOfBibitem
\bibitem[Saintillan and Shelley(2013)]{Saintillan2013}
D.~Saintillan and M.~J. Shelley, \emph{Comp. Rend. Physique}, 2013,
  \textbf{14}, 497--517\relax
\mciteBstWouldAddEndPuncttrue
\mciteSetBstMidEndSepPunct{\mcitedefaultmidpunct}
{\mcitedefaultendpunct}{\mcitedefaultseppunct}\relax
\EndOfBibitem
\bibitem[Gaffney \emph{et~al.}(2011)Gaffney, Gad{\^e}lha, Smith, Blake, and
  Kirkman-Brown]{Gaffney2011}
E.~A. Gaffney, H.~Gad{\^e}lha, D.~J. Smith, J.~R. Blake and J.~C.
  Kirkman-Brown, \emph{Annu. Rev. Fluid Mech.}, 2011, \textbf{43},
  501--528\relax
\mciteBstWouldAddEndPuncttrue
\mciteSetBstMidEndSepPunct{\mcitedefaultmidpunct}
{\mcitedefaultendpunct}{\mcitedefaultseppunct}\relax
\EndOfBibitem
\bibitem[Bhattacharjee \emph{et~al.}(2011)Bhattacharjee, Mc{D}onnell,
  Prabhakar, Yeo, and Friend]{Bhattacharjee2011}
P.~K. Bhattacharjee, A.~G. Mc{D}onnell, R.~Prabhakar, L.~Y. Yeo and J.~R.
  Friend, \emph{New J. Phys.}, 2011, \textbf{13}, 023005\relax
\mciteBstWouldAddEndPuncttrue
\mciteSetBstMidEndSepPunct{\mcitedefaultmidpunct}
{\mcitedefaultendpunct}{\mcitedefaultseppunct}\relax
\EndOfBibitem
\bibitem[Guillard and Ryther(1962)]{Guillard1962}
R.~R.~L. Guillard and J.~H. Ryther, \emph{Can. J. Microbiol.}, 1962,
  \textbf{8}, 229--239\relax
\mciteBstWouldAddEndPuncttrue
\mciteSetBstMidEndSepPunct{\mcitedefaultmidpunct}
{\mcitedefaultendpunct}{\mcitedefaultseppunct}\relax
\EndOfBibitem
\bibitem[Gibbs \emph{et~al.}(2011)Gibbs, Orta, Reddy, Koppers,
  Mart{\'{i}}nez-L{\'{o}}pez, de~la Vega-Beltr{\'{a}}n, Lo, Veldhuis, Jamsai,
  Mc{I}ntyre, Darszon, and O{'B}ryan]{Gibbs2011}
G.~M. Gibbs, G.~Orta, T.~Reddy, A.~J. Koppers, P.~Mart{\'{i}}nez-L{\'{o}}pez,
  J.~L. de~la Vega-Beltr{\'{a}}n, J.~C. Lo, N.~Veldhuis, D.~Jamsai,
  P.~Mc{I}ntyre, A.~Darszon and M.~K. O{'B}ryan, \emph{Proc. Natl. Acad. Sci.
  USA}, 2011, \textbf{108}, 7034--7039\relax
\mciteBstWouldAddEndPuncttrue
\mciteSetBstMidEndSepPunct{\mcitedefaultmidpunct}
{\mcitedefaultendpunct}{\mcitedefaultseppunct}\relax
\EndOfBibitem
\bibitem[McKinley and Sridhar(2002)]{McKinley2002}
G.~McKinley and T.~Sridhar, \emph{Annu. Rev. Fluid Mech.}, 2002, \textbf{34},
  375--415\relax
\mciteBstWouldAddEndPuncttrue
\mciteSetBstMidEndSepPunct{\mcitedefaultmidpunct}
{\mcitedefaultendpunct}{\mcitedefaultseppunct}\relax
\EndOfBibitem
\bibitem[Szabo(1997)]{Szabo1997}
P.~Szabo, \emph{Rheologica Acta}, 1997, \textbf{36}, 277--284\relax
\mciteBstWouldAddEndPuncttrue
\mciteSetBstMidEndSepPunct{\mcitedefaultmidpunct}
{\mcitedefaultendpunct}{\mcitedefaultseppunct}\relax
\EndOfBibitem
\bibitem[McKinley and Tripathi(2000)]{McKinley2000}
G.~McKinley and A.~Tripathi, \emph{J. Rheol.}, 2000, \textbf{44},
  653--670\relax
\mciteBstWouldAddEndPuncttrue
\mciteSetBstMidEndSepPunct{\mcitedefaultmidpunct}
{\mcitedefaultendpunct}{\mcitedefaultseppunct}\relax
\EndOfBibitem
\bibitem[Entov and Hinch(1997)]{Entov1997}
V.~Entov and E.~Hinch, \emph{J. Non-Newtonian Fluid Mech.}, 1997, \textbf{72},
  31--53\relax
\mciteBstWouldAddEndPuncttrue
\mciteSetBstMidEndSepPunct{\mcitedefaultmidpunct}
{\mcitedefaultendpunct}{\mcitedefaultseppunct}\relax
\EndOfBibitem
\bibitem[Chen \emph{et~al.}(2002)Chen, Notz, and Basaran]{Chen2002}
A.~U. Chen, P.~K. Notz and O.~A. Basaran, \emph{Phys. Rev. Lett.}, 2002,
  \textbf{88}, 174501\relax
\mciteBstWouldAddEndPuncttrue
\mciteSetBstMidEndSepPunct{\mcitedefaultmidpunct}
{\mcitedefaultendpunct}{\mcitedefaultseppunct}\relax
\EndOfBibitem
\bibitem[Eggers and F.(1994)]{Eggers1994}
J.~Eggers and D.~T. F., \emph{J. Fluid Mech.}, 1994, \textbf{262},
  205--221\relax
\mciteBstWouldAddEndPuncttrue
\mciteSetBstMidEndSepPunct{\mcitedefaultmidpunct}
{\mcitedefaultendpunct}{\mcitedefaultseppunct}\relax
\EndOfBibitem
\bibitem[Rodd \emph{et~al.}(2005)Rodd, Scott, and {Cooper-White}]{Rodd2005}
L.~Rodd, T.~P. Scott and G.~H. {Cooper-White}, J. J.;~Mc{K}inley, \emph{Appl.
  Rheol.}, 2005, \textbf{15}, 12--27\relax
\mciteBstWouldAddEndPuncttrue
\mciteSetBstMidEndSepPunct{\mcitedefaultmidpunct}
{\mcitedefaultendpunct}{\mcitedefaultseppunct}\relax
\EndOfBibitem
\bibitem[Tirtaatmadja \emph{et~al.}(2006)Tirtaatmadja, McKinley, and
  {Cooper-White}]{Tirtaatmadja2006}
V.~Tirtaatmadja, G.~H. McKinley and J.~J. {Cooper-White}, \emph{Phys. Fluids},
  2006, \textbf{18}, 043101\relax
\mciteBstWouldAddEndPuncttrue
\mciteSetBstMidEndSepPunct{\mcitedefaultmidpunct}
{\mcitedefaultendpunct}{\mcitedefaultseppunct}\relax
\EndOfBibitem
\bibitem[Eggers(1997)]{Eggers1993}
J.~Eggers, \emph{Phys. Rev. Lett.}, 1997, \textbf{71}, 3458–--3490\relax
\mciteBstWouldAddEndPuncttrue
\mciteSetBstMidEndSepPunct{\mcitedefaultmidpunct}
{\mcitedefaultendpunct}{\mcitedefaultseppunct}\relax
\EndOfBibitem
\bibitem[Eggers(1997)]{Eggers1997}
J.~Eggers, \emph{Rev. Mod. Phys.}, 1997, \textbf{69}, 865--930\relax
\mciteBstWouldAddEndPuncttrue
\mciteSetBstMidEndSepPunct{\mcitedefaultmidpunct}
{\mcitedefaultendpunct}{\mcitedefaultseppunct}\relax
\EndOfBibitem
\bibitem[Papageorgiou(1995)]{Papageorgiou1995}
D.~T. Papageorgiou, \emph{Phys. Fluids}, 1995, \textbf{7}, 1529--1544\relax
\mciteBstWouldAddEndPuncttrue
\mciteSetBstMidEndSepPunct{\mcitedefaultmidpunct}
{\mcitedefaultendpunct}{\mcitedefaultseppunct}\relax
\EndOfBibitem
\bibitem[Krieger and Dougherty(1959)]{Krieger1959}
I.~M. Krieger and T.~Dougherty, \emph{J. Rheol.}, 1959, \textbf{3},
  137--152\relax
\mciteBstWouldAddEndPuncttrue
\mciteSetBstMidEndSepPunct{\mcitedefaultmidpunct}
{\mcitedefaultendpunct}{\mcitedefaultseppunct}\relax
\EndOfBibitem
\bibitem[Ball and Richmond(1980)]{Ball1980}
R.~C. Ball and P.~Richmond, \emph{Phys. Chem. Liquids}, 1980, \textbf{9},
  99--116\relax
\mciteBstWouldAddEndPuncttrue
\mciteSetBstMidEndSepPunct{\mcitedefaultmidpunct}
{\mcitedefaultendpunct}{\mcitedefaultseppunct}\relax
\EndOfBibitem
\bibitem[Larson(1999)]{Larson1999}
R.~G. Larson, \emph{{The Structure and Rheology of Complex Fluids}}, Oxford
  University Press, Oxford, 1999\relax
\mciteBstWouldAddEndPuncttrue
\mciteSetBstMidEndSepPunct{\mcitedefaultmidpunct}
{\mcitedefaultendpunct}{\mcitedefaultseppunct}\relax
\EndOfBibitem
\bibitem[Hinch and Leal(1976)]{HinchLeal76}
E.~Hinch and L.~Leal, \emph{J. Fluid Mech.}, 1976, \textbf{76}, 187--208\relax
\mciteBstWouldAddEndPuncttrue
\mciteSetBstMidEndSepPunct{\mcitedefaultmidpunct}
{\mcitedefaultendpunct}{\mcitedefaultseppunct}\relax
\EndOfBibitem
\bibitem[Batchelor(1970)]{Batchelor1970}
G.~K. Batchelor, \emph{J. Fluid Mech.}, 1970, \textbf{44}, 419--440\relax
\mciteBstWouldAddEndPuncttrue
\mciteSetBstMidEndSepPunct{\mcitedefaultmidpunct}
{\mcitedefaultendpunct}{\mcitedefaultseppunct}\relax
\EndOfBibitem
\bibitem[ten Hagen \emph{et~al.}(2011)ten Hagen, van Teeffelen, and
  L{\"o}wen]{tenHagen2011}
B.~ten Hagen, S.~van Teeffelen and H.~L{\"o}wen, \emph{J. Phys.: Cond. Mat.},
  2011, \textbf{23}, 194119\relax
\mciteBstWouldAddEndPuncttrue
\mciteSetBstMidEndSepPunct{\mcitedefaultmidpunct}
{\mcitedefaultendpunct}{\mcitedefaultseppunct}\relax
\EndOfBibitem
\bibitem[Saragosti \emph{et~al.}(2012)Saragosti, Silberzan, and
  Buguin]{Saragosti2012}
J.~Saragosti, P.~Silberzan and A.~Buguin, \emph{PLoS ONE}, 2012, \textbf{7},
  e35412\relax
\mciteBstWouldAddEndPuncttrue
\mciteSetBstMidEndSepPunct{\mcitedefaultmidpunct}
{\mcitedefaultendpunct}{\mcitedefaultseppunct}\relax
\EndOfBibitem
\bibitem[Mortimer(1997)]{Mortimer1997}
S.~T. Mortimer, \emph{Human Reproduction Update}, 1997,  403--409\relax
\mciteBstWouldAddEndPuncttrue
\mciteSetBstMidEndSepPunct{\mcitedefaultmidpunct}
{\mcitedefaultendpunct}{\mcitedefaultseppunct}\relax
\EndOfBibitem
\bibitem[Qi \emph{et~al.}(2007)Qi, Moran, and Navarro]{Qi2007}
H.~Qi, M.~M. Moran and B.~Navarro, 2007,  1219--1223\relax
\mciteBstWouldAddEndPuncttrue
\mciteSetBstMidEndSepPunct{\mcitedefaultmidpunct}
{\mcitedefaultendpunct}{\mcitedefaultseppunct}\relax
\EndOfBibitem
\bibitem[Olson \emph{et~al.}(2011)Olson, Suarez, and Fauci]{Olson2011}
S.~D. Olson, S.~S. Suarez and L.~J. Fauci, \emph{J. Theor. Biol.}, 2011,
  \textbf{283}, 203--216\relax
\mciteBstWouldAddEndPuncttrue
\mciteSetBstMidEndSepPunct{\mcitedefaultmidpunct}
{\mcitedefaultendpunct}{\mcitedefaultseppunct}\relax
\EndOfBibitem
\bibitem[Doi and Edwards(1986)]{doiedw}
M.~Doi and S.~F. Edwards, \emph{{The Theory of Polymer Dynamics}}, {Oxford
  University Press}, 1986\relax
\mciteBstWouldAddEndPuncttrue
\mciteSetBstMidEndSepPunct{\mcitedefaultmidpunct}
{\mcitedefaultendpunct}{\mcitedefaultseppunct}\relax
\EndOfBibitem
\bibitem[Dunstan \emph{et~al.}(2012)Dunstan, Mi{\~n}o, Clement, and
  Soto]{Dunstan2012}
J.~Dunstan, G.~Mi{\~n}o, E.~Clement and R.~Soto, \emph{Phys. Fluids}, 2012,
  \textbf{24}, 011901\relax
\mciteBstWouldAddEndPuncttrue
\mciteSetBstMidEndSepPunct{\mcitedefaultmidpunct}
{\mcitedefaultendpunct}{\mcitedefaultseppunct}\relax
\EndOfBibitem
\bibitem[Bayly \emph{et~al.}(2011)Bayly, Lewis, Ranz, Okamoto, Pless, and
  Dutcher]{Bayly2011}
P.~V. Bayly, B.~L. Lewis, E.~C. Ranz, R.~J. Okamoto, R.~B. Pless and S.~K.
  Dutcher, \emph{Biophys J}, 2011, \textbf{100}, 2716--2725\relax
\mciteBstWouldAddEndPuncttrue
\mciteSetBstMidEndSepPunct{\mcitedefaultmidpunct}
{\mcitedefaultendpunct}{\mcitedefaultseppunct}\relax
\EndOfBibitem
\bibitem[Schmitz \emph{et~al.}(2000)Schmitz, Holcomb-Wygle, Oberski, and
  Lindemann]{Schmitz2000}
K.~A. Schmitz, D.~L. Holcomb-Wygle, D.~J. Oberski and C.~B. Lindemann,
  \emph{Biophysj}, 2000, \textbf{79}, 468--478\relax
\mciteBstWouldAddEndPuncttrue
\mciteSetBstMidEndSepPunct{\mcitedefaultmidpunct}
{\mcitedefaultendpunct}{\mcitedefaultseppunct}\relax
\EndOfBibitem
\bibitem[Bird \emph{et~al.}(1987)Bird, Curtiss, Armstrong, and Hassager]{dpl2}
R.~B. Bird, C.~F. Curtiss, R.~C. Armstrong and O.~Hassager, \emph{Dynamics of
  Polymeric Liquids}, Wiley-Interscience, New York, 2nd edn, 1987, vol. 2.
  {K}inetic theory\relax
\mciteBstWouldAddEndPuncttrue
\mciteSetBstMidEndSepPunct{\mcitedefaultmidpunct}
{\mcitedefaultendpunct}{\mcitedefaultseppunct}\relax
\EndOfBibitem
\end{mcitethebibliography}
\bibliographystyle{rsc} 
}

\end{document}